\documentstyle[psfig,epsf,conf-X,10pt]{article}
\begin{document} 
\small
\heading{
The X-ray properties of luminous infrared galaxies and their}
\heading{contribution to the X-ray background}

\par\medskip\noindent
\author{%
G. Risaliti$^{1}$
}
\address{
$^{1}$Dip. Astronomia, Largo E. Fermi 5, I-50125 Firenze, Italy;
}

\begin{abstract}
We present a study of the sample of luminous infrared galaxies (LIGs,
L$_{IR} >
10^{11} L_\odot$) observed in the hard (2-10 keV) X rays. The main
results are: 1)
most LIGs are powered both by AGN and starburst activity; 
2) the AGNs in
our sample are absorbed in the infrared by a lower N$_H$ than in the X-rays
or, alternatively, the dust-to-gas ratio is lower than galactic; 3)
the study of a subsample of sources observed in the 20-200
keV band indicates that most of the AGNs hosted by the LIGs are heavily
obscured
up to 100 keV and, therefore, their contribution to the X-ray
background
must be small.
\end{abstract}
\section{Introduction}
The Luminous Infrared Galaxies (LIGs) are a class of objects
characterized by high infrared luminosities (L$_{IR} > 10^{11}$ erg
s$^{-1}$). Both AGNs
and starbursts can power LIGs, but the relative
contribution of these two energy sources to the bolometric luminosity is
still unclear.

X-ray observations in the hard (2-10 keV) band can be a powerful tool
to unveil the AGN emission in the LIGs and to estimate its
contribution to the total luminosity. It is at present impossible to
analyze a representative LIGs sample in the 2-10 keV band, because the
hard X-ray observations performed up to now are strongly biased in
favor of
AGN--dominated sources. Nevertheless, a significant number of
IR-selected
LIGs with X-ray observations is now available both in the
literature and in public archives, and therefore a comparison between
the X-ray properties of
AGN--dominated and IR--selected LIGs is possible. We
collected the data of all the luminous infrared galaxies observed so
far in hard
X-rays and studied their X-ray emission and the correlation between
their
X-ray and infrared properties.

\section{X--ray and IR properties of the sample}

The X-ray properties of our sample of objects are very heterogeneous
both
in terms of brightness and spectral shape. Most of the objects optically
classified as type 1 Sys or QSOs are characterized by
bright X-ray emission (relative to the IR luminosity)
and their X-ray spectrum does not show indications for
significant cold absorption.  A significant fraction of the narrow line AGNs
are also
relatively bright in the X-rays and their spectrum is characterized by
a photoleletric cutoff ascribed to Compton thin absorbing gas
($\rm N_H < 10^{24}cm^{-2}$). The remaining objects optically classified
as AGNs are very weak in the X-rays.
This can be due to the intrinsic weakness of the AGN component or to an
absorbing column density higher than $10^{24}$cm$^{-2}$ (Compton thick
AGNs). Finally, a few objects are optically classified as starbursts and
are all very weak in the X rays. In Fig. 1 we plot the X/IR flux ratio
versus the infrared colour defined as C$_{IR}=2\times \frac{f_{25}}{f_{60}}$,
where f$_{25}$ and f$_{60}$ are the flux densities at 25 $\mu$m and at
60$\mu$m.
A clear correlation is apparent in Fig. 1: type 1 AGNs are
preferentially
in the high X/IR ratio and warm infrared colour part of the diagram.
Moving towards lower 25/60$\mu$m ratios we find lower X/IR ratios and an
increasing fraction of obscured AGNs at first, and of starbursts
afterwards.

\begin{figure}
\centerline{\mbox{\epsfxsize=9.0cm \epsffile{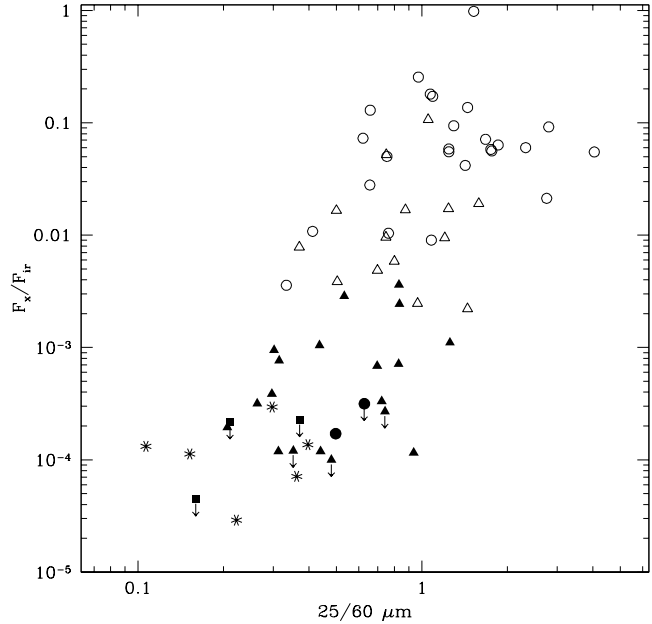}}}

\caption[] {X/IR flux ratio versus infrared colour for the sources of our
sample. Symbols: open circles: Seyfert 1s; open triangles:
Compton thin Seyfert 2s; filled triangles:
Compton thick Seyfert 2s; filled circles: broad absorption line quasars;
stars: starbursts.}
\end{figure}
A simple model in agreement with this correlation is shown in Fig. 2:
starting from a ``pure Seyfert 1'' point, at the top -right corner in
the plot, the oblique rightmost (light) line gives the expected location of
 AGN--dominated
objects, with an increasing X-ray absorbing column density moving
towards the bottom. The dotted
(dark) curves give different degrees of mixing of starburst and AGN, with the
AGN contribution lowering moving to the left.

\begin{figure}
\centerline{\mbox{\epsfxsize=7.0cm \epsffile{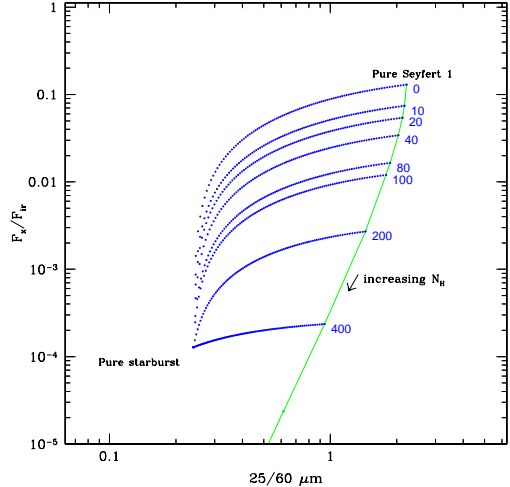}}}

\caption[] {Results of our model of mixed AGN and starburst contribution
to the X-ray and IR emission of LIGs. See text for details.}
\end{figure}
In this model
the absorption suffered by the IR radiation is only a small fraction of
that inferred from the X-ray column density by assuming a standard
dust--to--gas ratio. Indeed, we find that there is no way to explain the
correlation if we assume that (1) the same amount of material obscures both
the X-rays and the infrared and that (2) the dust--to--gas ratio is
galactic. This is because if none out of the two hypothesis above are
relaxed, the absorbed model gives a
pure--AGN curve that is almost horizontal, because the IR colour
decreases too fast with respect to the X/IR flux ratio.
A simple explanation for the difference between the absorption in the IR
and in the X-rays can be the following: if the density in the
circumnuclear torus decreases radially, the X--rays, emitted in the very
central region, are much more absorbed than the 25$\mu$m radiation,
emitted by the warm dust located along the inner face of the torus, far
from the plane of the accretion disk.

Besides the optical classification,
the simple picture depicted above is supported by additional pieces of 
evidence: 
1) broad lines in the polarized spectrum or in the near IR are found only in type 2 objects 
with warm IR colour, that, according to our model, are the AGN-dominated LIGs; 2) 
the large majority of the IR-cold
sources have steep (starburst-like) X-ray indices while
IR-warm sources have flatter (AGN-like) indices.
For a more detailed discussion about these
issues, we remind to a forthcoming paper \cite{Rprep}.

\section {The 20-200 keV emission of LIGs}
Our sample includes a subsample of 13 sources optically classified as
type 2 AGNs and observed by BeppoSAX up to
200 keV. 6 out of the 11 sources that are
Compton--thick in the 2-10 keV range are completely absorbed also
in the 10 to 200 keV band, therefore
 implying a column density N$_H > 10^{25}$ cm$^{-2}$.
Only two of the 11 Compton thick sources have an excess in the 15-100
keV range, while for the remaining three the hard 15-100 keV X-ray emission
is unconstrained. The shortage of objects with 10$^{24}$cm$^{-2} < $N$_H <
10^{25}$ cm$^{-2}$ already pointed out in a sample of optically
selected Seyfert 2s \cite{RMS}, would have
important consequences in the synthesis models of the X-ray
background, since this class of Sy2 should contribute significantly to
the XRB in the 10 keV to 200 keV band, where most of the XRB energy is
emitted.
These models are by and large successful in synthesizing the XRB from
the contributions of individual AGNs; they must however include a
dominant contribution from absorbed, type 2 AGNs, which up to now have
been observed only at low redshifts and low luminosities. Assuming that
type 1 and type 2 AGNs evolve in the same way (as predicted by unified
models), the zero-th order extrapolation has to face several
discrepancies, and could be cured only by adding extra type 2 sources at
intermediate or high redshifts \cite{GRS}. It has been proposed that
LIGs could be the required additional sources, thanks to their high
luminosity and high spatial density. However, our work indicates that
even those sources in which the presence of a luminous AGN is strongly
supported by several optical and IR indicators, are on average 
very dim in the hard
X-rays up to 100 keV, because of heavy obscuration. Therefore, the
contribution of this class of sources to the XRB is negligible.

\begin{acknowledgements}
Much of this work has been don in collaboration with R. Gilli, R.
Maiolino and M. Salvati.
The author acknowledges the partial financial support
from the Italian Space Agency (ASI)
through the grant ARS--99--15.
\end{acknowledgements}

\vfill
\end{document}